\documentclass[10pt,journal]{IEEEtranTCOM}
\usepackage[english]{babel}
\usepackage[usenames]{color}
\usepackage[cp1250]{inputenc}
\usepackage{amsfonts}
\usepackage{amsthm}
\usepackage{graphicx}
\usepackage{epsfig}
\usepackage{mathrsfs}
\usepackage{amsmath}
\usepackage{algorithm}
\usepackage{algorithmic}

\theoremstyle{plain}

\begin{document}
\newcommand{\bea}{\begin{eqnarray}}
\newcommand{\eea}{\end{eqnarray}}
\newcommand{\be}{\begin{equation}}
\newcommand{\ee}{\end{equation}}
\newcommand{\beas}{\begin{eqnarray*}}
\newcommand{\eeas}{\end{eqnarray*}}
\newcommand{\bs}{\backslash}
\newcommand{\bc}{\begin{center}}
\newcommand{\ec}{\end{center}}
\def\SC {\mathscr{C}}

\title{Normalized rotation shape descriptors \\ and lossy compression of molecular shape}
\author{\IEEEauthorblockN{Jarek Duda}\\
\IEEEauthorblockA{Jagiellonian University,
Golebia 24, 31-007 Krakow, Poland,
Email: \emph{dudajar@gmail.com}}}
\maketitle

\begin{abstract}
There is a common need to search of molecular databases for compounds resembling some shape, what suggests having similar biological activity while searching for new drugs.
The large size of the databases requires fast methods for such initial screening, for example based on feature vectors constructed to fulfill the requirement that similar molecules should correspond to close vectors.
Ultrafast Shape Recognition (USR) is a popular approach of this type. It uses vectors of 12 real number as 3 first moments of distances from 4 emphasized points. These coordinates might contain unnecessary correlations and does not allow to reconstruct the approximated shape. In contrast, spherical harmonic (SH) decomposition uses orthogonal coordinates, suggesting their independence and so lager informational content of the feature vector. There is usually considered rotationally invariant SH descriptors, what means discarding of some essential information.

In contrast, this article discusses framework for descriptors with normalized rotation, for example by using principal component analysis (PCA-SH).
As one of the most interesting are ligands which have to slide into a protein like a key, we will introduce descriptors optimized for such flat elongated shapes. Bent deformed cylinder (BDC) describes the molecule as a cylinder which was first bent, then deformed such that its cross-sections became ellipses of evolving shape. Legendre polynomials are used to describe the central axis of such bent cylinder, which will be referred as its spine. Additional polynomials are used to define evolution of such elliptic cross-section along the main axis. There will be also discussed bent cylindrical harmonics (BCH), which uses cross-sections described by cylindrical harmonics instead of ellipses. All these normalized rotation descriptors allow to reconstruct (decode) the approximated representation of the shape, hence can be also used for lossy compression purposes.
\end{abstract}
\section{Introduction}
\begin{figure}[t!]
    \centering
        \includegraphics[width=8cm]{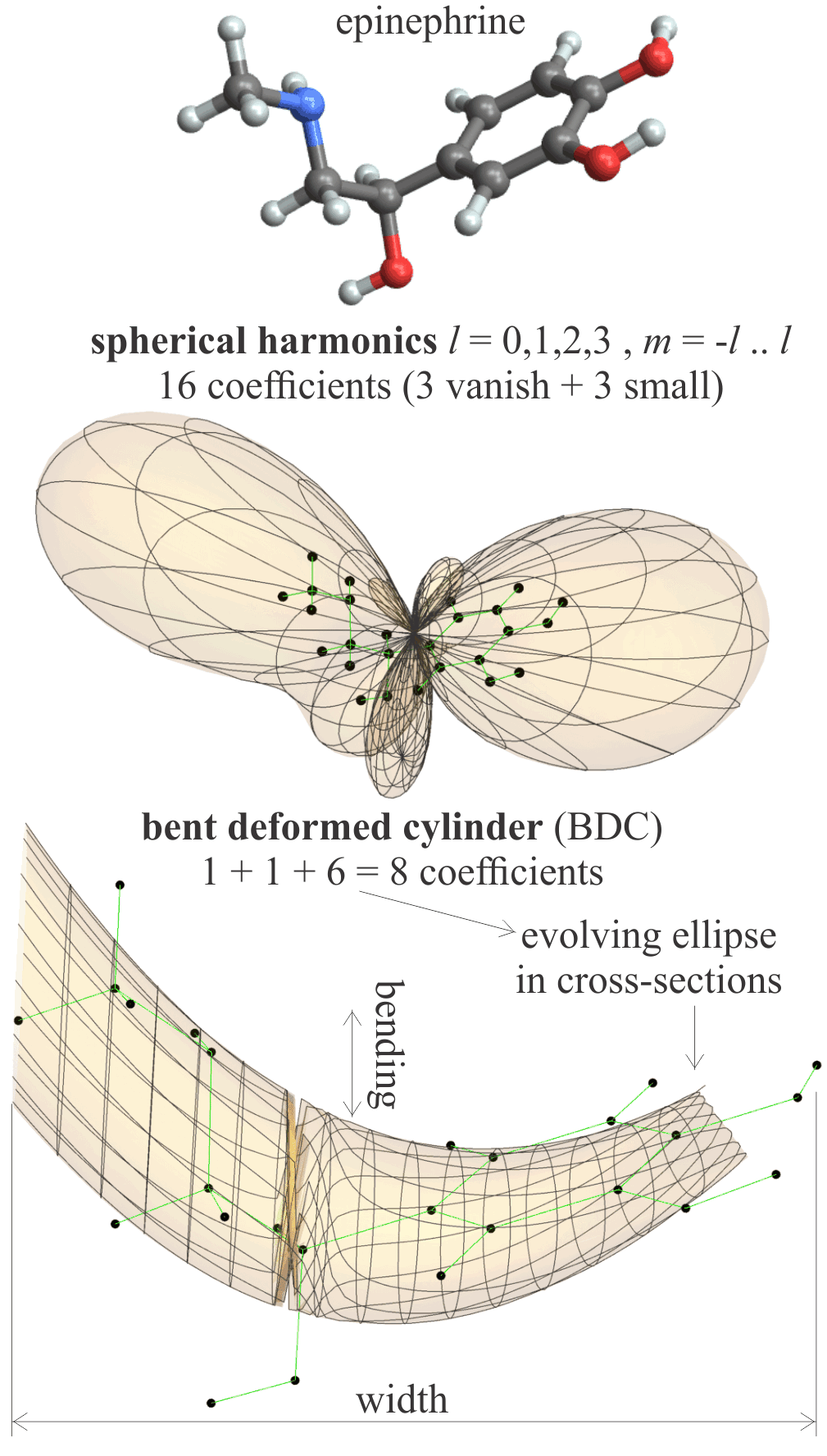}
        \caption{Examples of decoded approximated shape for descriptors of epinephrine. For spherical harmonics around the centroid, the 3 coordinates for $l=1$ are nearly vanishing. Additionally the PCA choice of rotation make 3 of 5 coefficients for $l=2$ small (but not negligible). The BDC descriptor seems to better correspond to the shape - it describes the general bending of the molecule, and evolution from nearly circular to flattened on the carbon ring. 6 of 8 coefficients describe evolution of shape of ellipse in the cross-sections. }
        \label{intr}
\end{figure}
Early stages of modern drug devolvement usually consist of virtual screening of database of available molecules before the expensive lab testing. This virtual screening usually base on similarity of shape of already known active molecules and may consist of multiple stages, as evaluation of similarity is a complex and not well-posed problem.

There are known costly accurate methods which take two molecules and evaluate their similarity, for example trying to align them~\cite{align} or based on graph kernels~\cite{kernel}. As the databases contain millions of entries, practical situations require some initial screening before such pairwise comparison of the most promising candidates.

For performance reasons, we would like to use vectors of features for such initial screening~\cite{finger}, which will be referred as \emph{fingerprints}, such that similar molecules are expected to correspond to vectors close in some metric. It allows to replace the costly evaluation of similarity with inexpensive metric for a relatively short vector. Additionally, it allows for clustering in the multidimensional space of features to obtain classification into groups of similar molecules. Some fingerprints may allow to reconstruct the approximated shape - we will refer to them as \emph{descriptors}. \\

There is a difficult question how to describe a complex molecule as a set of features - for performance reasons we will focus on using a fixed length vectors of real numbers (e.g. 12), like in examples in Fig. \ref{intr} which will be discussed further.

Ultrafast Shape Recognition (USR)~\cite{USR} is a popular example of such fingerprints. It uses feature vectors of 12 numbers as first three moments for distances from four vertices: the centroid (ctd), the closest atom to ctd (cst), the farthest atom to cst (fct) and the farthest atom to fct (ftf).

\begin{figure}[t!]
    \centering
        \includegraphics[width=8cm]{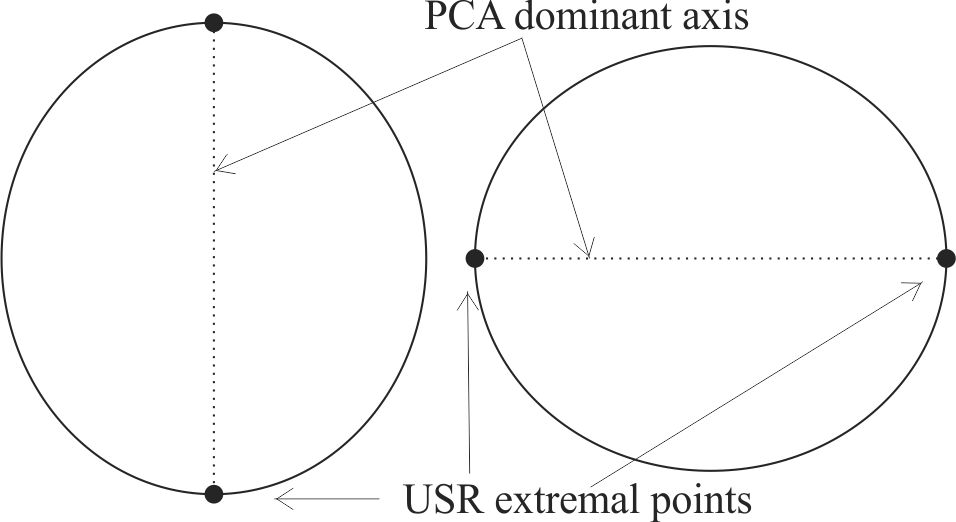}
        \caption{Example of issues with continuity - for a nearly spherical object, small deformation can change the emphasized axis or points used to anchor the fingerprint. It can be handled by modifying behavior in such rare nearly symmetric situations: for example finding descriptors for both choices and use some their average - at cost of weakening the decodability property. }
        \label{cont}
\end{figure}

Before discussing further examples, let us try to verbalize properties we would like for a perfect descriptor of a shape (or chemical properties) of a molecule:
\begin{enumerate}
  \item \emph{representativeness} - close fingerprints should correspond to similar molecules, or in other words: distant molecules to distant fingerprints,
  \item \emph{continuity} - small perturbation of molecule should not lead to a large change of its fingerprint,
  \item \emph{selectiveness} - for performance reasons we would like the vectors to be as short as possible, maximally exploit all the used coefficients, so they should represent only the most significant features which might be meaningful for the interesting process like ligand bonding,
  \item \emph{independence} - analogously, the coefficient should not be correlated,
  \item \emph{decodability} - the fingerprint should allow to reconstruct the used approximation of shape (be a descriptor), can be treated as its lossy compression,
  \item \emph{faithfulness} - if decodable, the approximation should agree with essential qualitative and quantitative properties of the molecule, should not introduce artifacts.
\end{enumerate}
The 1) and 2) are necessary to make the fingerprint metric in agreement with similarity of molecules. Unfortunately, USR and descriptors based on normalized rotation may have issues with the continuity, like in example in Fig. \ref{cont}. However, these are relatively rare symmetric cases, for example ligands which are of the main interest here are usually elongated. There will be also discussed ways to eventually repair this issue if needed.

The 3) and 4) regard maximization of descriptive potential of fingerprints of a given size. It can be realized by first producing a larger vector of features, then use the principal component analysis (PCA) to find their linear combinations which will be independent and the most significant. By using a base of orthogonal functions like spherical harmonics,  we can try to design descriptors directly fulfilling 3), 4) this way. However, given type of data may still contain some hidden correlations, for example of ligand size and shape - it might be beneficial to optimize the final descriptors by the using the PCA procedure.

Finally 5) and 6) regard the possibility of reconstruction (decoding) of the approximated shape, which direct application is lossy compression for example of a large set of molecules and their confirmations. Additional advantage is the possibility of evaluating correspondence of the represented approximation and the actual molecule. As this correspondence is difficult to quantitatively formalize, possibility of decoding the represented approximated shape allows for example to visually evaluate faithfulness of some representation for a given type of molecules.

Let us look at USR from the perspective of these 6 properties. There is no guarantee of 1): that changes of shape have to modify the moments. For example if the four points are in line (they often nearly are), rotation of single atoms around this line would not change the moments. As in example in Fig. \ref{cont}, continuity might be violated for nearly symmetric objects. It has no guarantee of selectiveness - that these 12 numbers corresponds to the most essential features. It does not fulfill independence neither: two of the points (ctd and cst) can be close to each other, making moments from them strongly correlated. There is also no direct way to reconstruct approximated shape from USR coefficients - 5) and 6) are also not fulfilled.\\

In contrast to USR, we will focus here on descriptors - which directly represent some approximated shape, what suggests a better correspondence and control. The standard approach is to use real spherical harmonics (SH)~\cite{harm}, describing radius in all spherical directions from a chosen point (usually centroid) of some surface envelope. As such description depends on the global rotation, the rotation independent coefficients are often used, which average the original coefficient - discarding potentially valuable information (selectiveness).

In contrast, we will focus on normalized rotation descriptors to be able to use all information from the low order coefficients (instead of their averages), which usually carry more crucial information. One approach of such normalization of rotation is to use PCA base, getting PCA-SH already used for a general purpose shape descriptors~\cite{PCAharm}. Its cost is a possibility of loosing continuity, as illustrated in Fig. \ref{cont}. However, these are relatively rare symmetric situations, especially for ligands which are usually elongated. There will be also discussed a way to completely remove this kind of issues, at cost of reduced decodability (not directly important for virtual screening purposes).

The SH approach can only express shapes (surface envelopes) having a specific property: that for any spherical direction from the central point, there is exactly one point of the surface. This property is weaker than convexity, for example fulfilled for '$\vee$' shape, however it is not fulfilled for '$\cup$' shape. Moreover, the shapes of low order spherical harmonics, presented in Fig. \ref{sh}, might not correspond well to the actual shape of molecule, leading to artifacts as in Fig. \ref{intr}.

Hence, there will by introduced descriptors which should be more appropriate for probably the most interesting molecules here: ligands. They are usually elongated to slide into a target protein, what suggests to use description around the longest axis, for example the dominant eigenvector of PCA. The molecule can be generally bent for example in '$\cup$' or '$\sim$' type of shape, hence we will describe the situation around some curve along this central axis - this curve will be referred as spine of the molecule and approximated using Legandre polynomials. Finally, there will be discussed two approaches to describe situation around this spine: bent deformed cylinder (BDC) where local situation is described as an ellipse, and bent cylindrical harmonics (BCH) where local situation is described by cylindical harmonics.

The presented framework is very general and its options and parameters should be optimized while designing descriptors for a specific application, basing on types of molecules and preferred evaluation of similarity, for example given by a reference metric which might be costly.\\

While real molecule can change its conformation (shape), all discussed descriptors assume a fixed conformation. A fixed length descriptors we assume here for performance reasons, has rather insufficient capabilities for describing all conformations. Hence, the standard approach is to generate all possible confirmations using some angle quantization, for example 15 degrees, and generate descriptor for each of them, treating them independently. Alternative approach is to encode both the average coefficients for all such conformations, and additionally their variance in this ensemble of conformations to describe its conformation mobility.

This is one of reasons for requiring low cost of calculation of the descriptor. Hence, we will focus on representing atoms as points instead of for example electron density, what allows to replace integral with sums. Generally, each of $n$ atoms can be assigned some weight: $w_i:\ i=1,\ldots,n$ which will be used for averages, as importance for approximations. For simplicity we focus on the $w=1$ case, making e.g. average position being the centroid. Alternatively, choosing this weight as atomic mass of given atom, what would greatly reduce significance of hydrogens, would make the average position the center of mass (barycenter).

The evaluation of such descriptor is a far nontrivial task, for example minimization of mean square error might lead to a significant over-fitting, making the decoded shape very different from the expected one. Another reason for required caution while trying to minimize some evaluation metric is the continuity: if such a metric would have multiple minima, a small perturbation could lead to a different one, and so a different descriptor.

Beside shape, other properties like electronnegativity are also crucial for biological activity, and can be simply expressed by a few numbers, which can be included in the feature vector. For example as coefficients of polynomial approximating their evolution along the main axis of molecule.

\section{Spherical harmonics}
We will now discuss the spherical harmonics, their PCA normalization of rotation and handling the continuity issues.
\subsection{General theory}
\begin{figure}[t!]
    \centering
        \includegraphics[width=8cm]{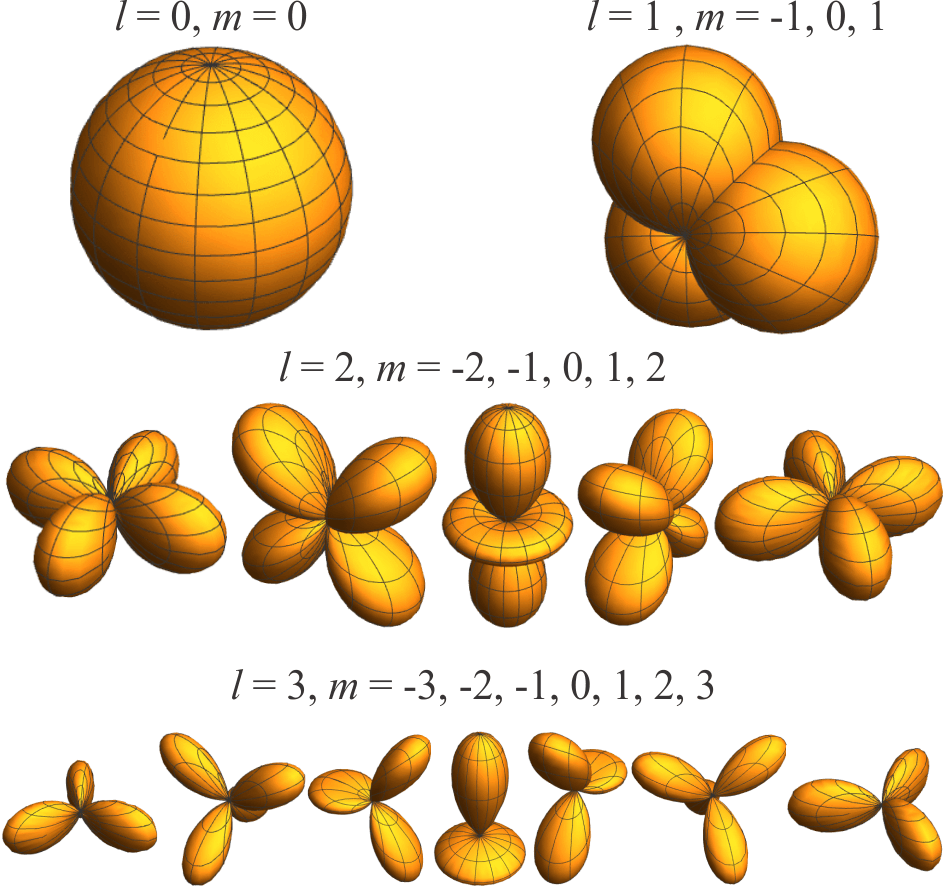}
        \caption{Spherical plots of spherical harmonics for $l=0,1,2,3$ and $m=-l, ..., l$. For $l=0$ it is the central sphere - the corresponding coordinate represents average distance from the central point. For $l=1$ they are 3 spheres in $x$, $y$, $z$ directions, which coefficients nearly vanish if the central point is the centroid. }
        \label{sh}
\end{figure}

Real spherical harmonics are a complete orthonormal base useful for approximation of the distance in all direction from a fixed point of a surface envelope:
\be r(\theta,\varphi)=\sum_{l=0}^L \sum_{m=-l}^l a_{lm} y_{lm}(\theta,\varphi) \label{sheq}\ee
where we assume standard spherical coordinates:
$$x=r \sin\theta \cos \varphi,\ y=r \sin \theta \sin \varphi,\ z=r\cos \theta$$
Figure \ref{sh} illustrates them for $l=0,1,2,3$. The formulas can be expressed using the $x,\ y,\ z$ coordinates, for example a few first are (omitted normalization terms):
$$y_{00}\propto 1,\quad y_{1,-1}\propto\frac{y}{r},\quad y_{10}\propto\frac{z}{r},\quad y_{11}\propto\frac{x}{r},$$
$$y_{2,-2}\propto\frac{xy}{r^2},\quad y_{2,-1}\propto\frac{yz}{r^2},\quad y_{20}\propto\frac{-x^2-y^2+2z^2}{r^2},$$
$$y_{21}\propto\frac{zx}{r^2},\quad y_{22}\propto\frac{x^2-y^2}{r^2}$$

The $a_{lm}$ coefficients depend on the global rotation - rotating the system transforms them accordingly to the corresponding Wigner rotation matrices $R^l$:
$$a'_{lm} = \sum_{m'=-l}^l R^l_{mm'} a_{lm'}$$

Having such representation of two surface envelopes ($A$ and $B$), we should rotate one of them ($B$) to minimize the mean square error (MSE):
$$MSE=\int(r_A-r_B)^2d\Omega=\sum_l \sum_m (a_{lm}-b'_{lm})^2$$
This minimization corresponding to finding the best alinement is costly. It can be be greatly simplified by using rotationally invariant fingerprints (RIFs) instead:
$$A_l=\sqrt{\sum_{m=-l}^l a^2_{lm}}$$
However, such averaging discards potentially essential information about the shape, as the low order coefficients are usually the most significant. Additionally, it loses the decodability property, as the averaged coefficients do not allow to retrieve the $r$ function. Hence, we will normalize the rotation before fitting coefficients to be able to use all the low order coefficients.
\begin{figure*}[t!]
    \centering
        \includegraphics[width=16cm]{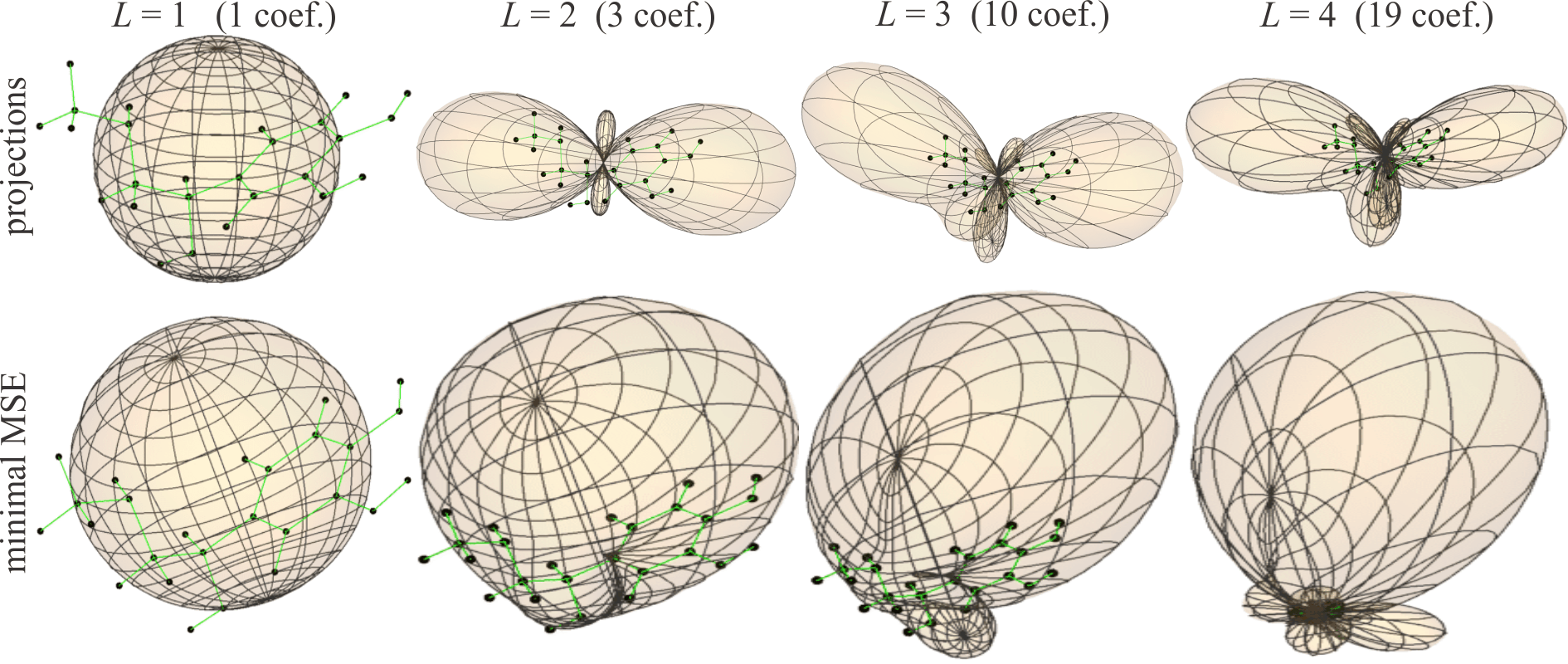}
        \caption{PCA-SH approximations of epinephrine using direct projection (\ref{proj}) and minimization of mean square error (MSE). The "coef." is the number of significant coefficients. As we can see, the latter leads to serious overfitting issues - the approximation is "swelling" in directions not covered by atoms. It shows difficulty of evaluating the resemblance, minimal MSE does not correspond to the best agreement. }
        \label{shepi}
\end{figure*}
\subsection{Normalization of translation and rotation} \label{normal}
For each molecule we would like to choose a unique position (translation) and rotation which will be later used to generate descriptors (or can be used to find an alignment).
This choice should be continuous: small changes of shape should lead to similar normalization. However, as illustrated in Fig. \ref{cont}, nearly symmetric situations could lead to discontinuity, what will be discussed later.

As mentioned, we will represent $n$ atoms as point with original coordinates $\textbf{v}_i=\{v_{i1},v_{i2},v_{i3}\}\equiv \{x_i,y_i,z_i\}$ for $i=1,\ldots,n$. We need first to emphasize some central point, the natural choice is to use the centroid and to shift the center of coordinates there, reassigning:
$$\textbf{v}'_i\ =\  \textbf{v}_i-\frac{1}{n} \sum_{j=1}^n \textbf{v}_j $$

Now we need to choose an unambiguous normalized rotation - a natural choice is to use PCA for this purpose. $C_{ab}=\sum_i v'_{ia}\cdot v'_{ib}$ is $3\times 3$ covariance matrix. Denote by $e_k$ its sorted eigenvectors: $C \textbf{e}_k = \lambda_k \textbf{e}_k$ for $\lambda_1\geq \lambda_2\geq \lambda_3$. We will use $\{\textbf{e}_k\}$ to define the normalized rotation.

Observe that $-\textbf{e}_k$ is also a valid eigenvector: we need to choose their signs in a unique way. There are many ways to do it, the applied one is to look at the minimal and maximal ($min_k,\ max_k$) value of the projection: $\{\textbf{v}'_i\cdot \textbf{e}_k:\ i=1,\ldots,n\}$. Usually $-min_k\neq max_k$, let us arbitrarily choose that we want $-min_k < max_k$, changing the sign of eigenvector if it is not true: $\textbf{e}_k = -\textbf{e}_k$. The $-min_k < max_k$ condition intuitively means that the molecule is denser on the negative side: if we would hang it in the middle between $min_k$ and $max_k$, the left hand side part would be heavier.

Above choice of sign should be made for $\textbf{e}_1$ and $\textbf{e}_2$. The sign of $\textbf{e}_3$ should be chosen to preserve orientation:
$$ \textrm{if}\quad det[\textbf{e}_1 \textbf{e}_2 \textbf{e}_3]<0\quad \textrm{then}\quad \textbf{e}_3=-\textbf{e}_3$$
Otherwise we would transform the molecule into its enantiomer. Finally we can perform the normalization - transform the points to the eignebase:
\be v_{ik}= \textbf{v}'_i\cdot \textbf{e}_k\ee

To summarize, the normalization consists of the following steps:
\begin{enumerate}
  \item move the center of coordinates to the centroid,
  \item calculate covariance matrix and its sorted eigenvectors $\textbf{e}_k$,
  \item change signs of $\textbf{e}_1$ and $\textbf{e}_2$ if needed to ensure $-min_k < max_k$,
  \item change sing of $\textbf{e}_3$ if needed to ensure proper orientation,
  \item transform points to the new base.
\end{enumerate}

\subsection{Fitting coefficients} \label{fitting}
The spherical harmonics are orthonormal for the
\be (f,g):=\int f(\theta, \varphi) g(\theta,\varphi) d\Omega\ee
scalar product. Hence the coefficients to minimize the mean square error are $a_{lm}=(r,y_{lm})$. However, for performance reason we would like to avoid this integration, representing atoms as just points. We can use approximated scalar product for this purpose:
\be [f,g]:=\sum_{i=1}^n f(\theta_i,\varphi_i) g(\theta_i,\varphi_i)\ee
where $(r_i,\theta_i,\varphi_i)$ are spherical coordinates of $i$-th point (atom). Now approximated coefficients are
\be a_{lm}=\frac{[r,y_{lm}]}{[y_{lm},y_{lm}]}=\frac{\sum_i r_i y_{lm}(\theta_i,\varphi_i)}{\sum_i \left(y_{lm}(\theta_i,\varphi_i)\right)^2} \label{proj}\ee
The original base of spherical harmonics is no longer normalized for the $[\cdot,\cdot]$ scalar product, hence the projection requires $[y_{lm},y_{lm}]$ denominator. Additionally, as we also lose orthogonality, this projection still does not ensure minimization of MSE. Instead of projection, we can find $a_{lm}$ coefficients minimizing the MSE: $\sum_i \left(r_i - \sum_{lm} a_{lm} y_{lm}(\theta_i,\varphi_i)\right)^2$.

However, as shown in Fig. \ref{shepi}, this can lead to serious overfitting issues: points enforce constraints only on some $(\theta_i,\varphi_i)$ directions. The remaining spherical directions can choose any values to minimize error for the constrained directions. It can cause "swelling" in unconstrained directions, completely loosing correspondence with the shape of molecule. This pathological behavior could be reduced by minimizing also coefficients, for example by adding $\sum_{lm} (a_{lm})^2$ to the minimized functional.

Surprisingly, as we can see in Fig. \ref{shepi}, the approximated approach of just using projections leads to much better shape agreement here. It shows the difficulty of even evaluating the resemblance of shapes.
\subsection{Handling discontinuities}
Unfortunately, the presented normalization of rotation has potential sources of discontinuity: small perturbation might change the order of eigenvectors or their sign, leading to a potentially very different description. It can happen near symmetric situations: when two eigenvalues are nearly equal, or $-min\approx max$, what seems rare for usually flat and elongated ligands.

This issue can be ultimately repaired at cost of decodability. Specifically, for this purpose the encoder should first test such closeness to a dangerous symmetric situation, e.g. $\lambda_1-\lambda_2 < \epsilon$ for some chosen parameter $\epsilon$. In this case, it should find the descriptors for all possible close alternative rotations, for example with switched $\textbf{e}_1$ and $\textbf{e}_2$ if $\lambda_1 \approx \lambda_2$. Then use a linear combination of these descriptors as the final descriptor:
\be \frac{\textbf{d}+\textbf{d}'}{2}+\frac{\lambda_1-\lambda_2}{\epsilon} \frac{\textbf{d}-\textbf{d}'}{2} \ee
where $\textbf{d}$ is the original descriptor, $\textbf{d}'$ is for switched $\textbf{e}_1$ and $\textbf{e}_2$. This way we have their average is $\lambda_1=\lambda_2$ and $\textbf{d}$ completely dominates for $\lambda_1-\lambda_2=\epsilon$.

This approach smoothen discontinuity. The cost, beside additional calculations, is the decodability: such linear combination of descriptors do not longer corresponds to an actual shape. However, the requirement of continuity and decodability usually appear separately: the former in virtual screening where we do not need to decode, the latter in lossy compression, where discontinuity is not a problem.
\subsection{Discarding some coordinates}
As the number of used coefficient affects the cost of both storing and comparing fingerprints, we would like to discard the less meaningful coordinates. We have 6 parameters while choosing position and rotation - we should be able to use them to enforce vanishing of 6 of $a_{lm}$ coefficients for optimization purposes. The discussed normalization nearly fulfills this requirement:  $a_{1,-1}$, $a_{1,0}$, $a_{1,1}$ are nearly vanishing, $a_{2,-2}$, $a_{2,-1}$, $a_{2,1}$ are small: they can be omitted while designing a descriptor.

The choice of centroid as the center of coordinates makes the $l=1$ nearly vanish. For example $y_{1,1}\propto \frac{x}{r}$, hence its average is nearly zero. In practice these coefficients turn out smaller than $10^{-10}\cdot a_{00}$.

The additional choice of 3 degrees of freedom for rotation should allow to cancel 3 of 5 from $l=2$ coefficients. The simplest are for $m=-2,-1,1$: correspondingly $\propto \frac{xy}{r^2}$, $\frac{yz}{r^2}$ and $\frac{xz}{r^2}$. The PCA coordinates approximate the shape as ellipsoid. Let us check that for an ellipse with radii $a$ and $b$, the directions zeroing average $xy$ are its axes:

$$\int_{-b}^b\int_{-a\sqrt{1-(y/b)^2}}^{-a\sqrt{1-(y/b)^2}} x'y' dx dy=\frac{\pi}{8}ab(a^2-b^2)\sin(2\alpha)$$

for $x'=x \cos(\alpha)-y \sin(\alpha)$, $y'=x\sin(\alpha)+y\cos(\alpha)$ rotated coordinates. From symmetry, average $xy$ is always zero for a circle ($a=b$). For an ellipse, it is zeroed only for both axes ($\sin(2\alpha)=0$). Numerical integration suggests that it also true for the interesting $\frac{xy}{r^2}$ function.

These are only suggestions that $a_{2,-2},\ a_{2,-1},\ a_{21}$ should be small for PCA normalization of rotation and so can be safely removed from the descriptor. In experimental tests these coefficients are usually below $0.1\cdot a_{00}$, however there are also exceptions. One could try to modify the PCA rotation to minimize these coefficients, however it is an additional cost and carries risk of weakening continuity, as there is no guarantee that there is only a single such rotation: a small perturbation could lead to a different normalization and so description.

\section{Bent deformed cylinder and cylindrical harmonics}
As we could see in Fig. \ref{shepi}, the shape represented by fitted spherical harmonics might contain some artifacts due to the characteristic shapes of SH as presented in Fig. \ref{sh}. Hence, we would like to design descriptors optimized for types of shapes appearing in a given problem. In the looking most applicable situation of virtual screening, the interesting molecules (ligands) need to fit into a protein, hence they often have elongated and flattened shape. It suggests to start with approximating it as a bent rod, which will be referred as spine, then describe the behavior of surface envelope around it.
\subsection{Normalization}
Let us start with normalization as discussed in Section \ref{normal} for the spherical harmonics case: first shift the center of coordinates to the centroid, then rotate it accordingly to the eigenvectors of the covariance matrix (PCA), with unambiguous choice of signs.

The dominant eigenvector, which corresponds to the $x$ axis now, is the main axis along the (potentially) elongated molecule - we will use it as the base of our description. For convenience, let us normalize the $x$ current coordinates to make them use the $[-1,1]$ range - reassign:
$$x_i = a x_i + b,\ \textrm{where}\ a=\frac{2}{max-min},\ b=1-a\cdot max$$
and $min$, $max$ are minimal and maximal values of the original $\{x_i\}$. Observe that this linear transformation: making that $x$ cover $\in[-1,1]$ range, makes that the center of coordinate is no longer the centroid. The $width=max-min$ is the scaling factor which should be used as one of coefficients of our descriptor.

\begin{figure}[t!]
    \centering
        \includegraphics[width=8cm]{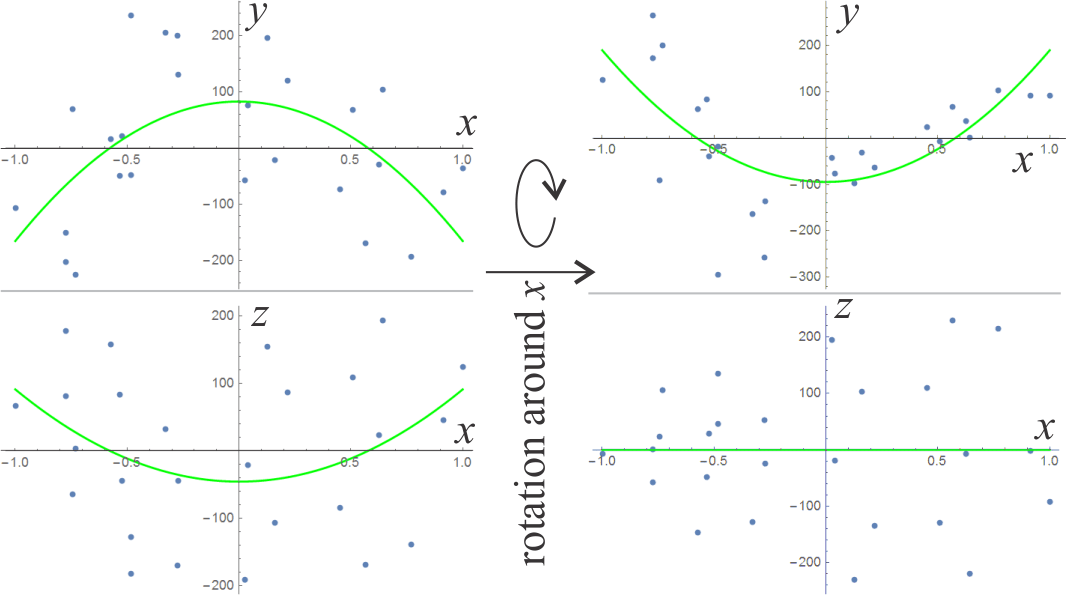}
        \caption{Normalization of rotation around $x$ axis to cancel the bending coefficient (of $3x^2-1$) in $z$ axis and make it positive for $y$ axis. Hence the molecule has always 'u'-shape in $xy$ and is approximately flat in $xz$.}
        \label{spine}
\end{figure}

The next subsection discusses additional step of normalization: rotate around $x$ axis to make the lowest order of bending zero in $z$ direction, and $\cup$-like in $y$ direction as presented in Fig. \ref{spine}, what allows to discard one coefficient (bending in $z$ direction).

\subsection{Fitting the spine with Legende polynomials}
\begin{figure}[t!]
    \centering
        \includegraphics[width=8cm]{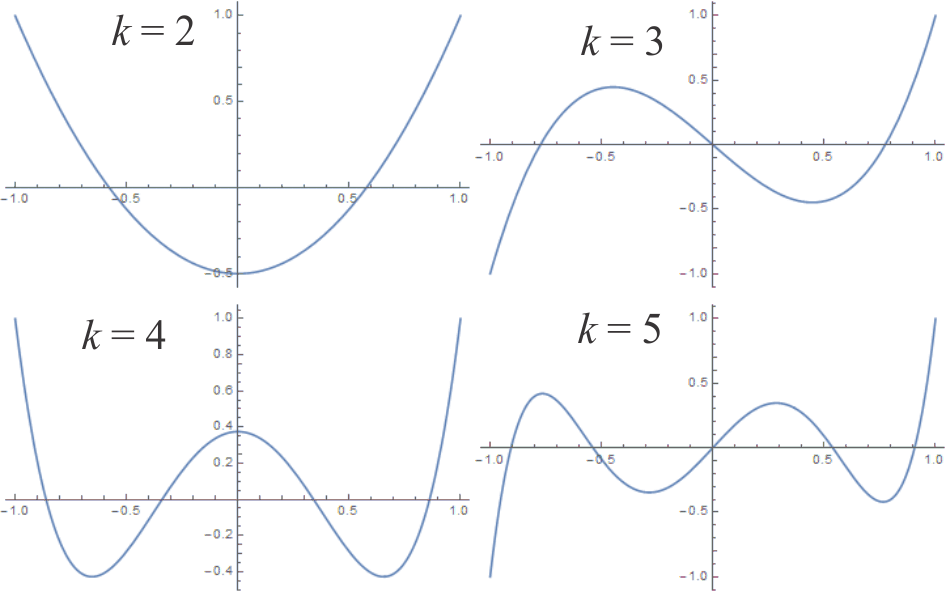}
        \caption{Some Legandre polynomials $P_2(x)=(3x^2-1)/2$, $P_3(x)=(5x^3-3x)/2$, $P_4(x)=(35x^4-30x^2+1)/8$, $P_5(x)=(63x^5-70x^3+15x)/8$ for interpolating the spine: the dominant axis of PCA is the horizontal $x$ axis for all of them.}
        \label{leg}
\end{figure}

We will now find the description of bent central rod of our descriptor, which will be referred as spine of the molecule.
The natural approach is to approximate it with polynomials. Our choice of the main axis ($x$) restricts the space of polynomials. It was obtained from PCA, hence it minimizes $\sum_i (y_i)^2$ and $\sum_i (z_i)^2$. As the spine should also minimize these distances, using $p=a$ or $p=ax$ polynomials does not make any sense as it would just shift or rotate our PCA main axis. Hence, we should use base of polynomials which are orthogonal to $1$ and $x$ polynomials - for the natural $(f,g)=\int_{-1}^1 f(x)g(x) dx$ scalar product, these are Legandre polynomials. A few of them are illustrated in Fig. \ref{leg}. Performing PCA on their graphs in $[-1,1]$ range would lead to $x$ as the main axis.

As discussed in Section \ref{fitting}, due to using point representation, we can find the coefficient using projections for $[f,g]=\sum_i f(x_i) g(x_i)$ approximated scalar product:
\be c^y_k = \left(\sum_i y_i P_k(x_i)\right) / \left(\sum_i (P_k(x_i))^2 \right)\ee
and analogously for $c^y_k$. As previously, one could minimize MSE instead of using these projections, however, it does not have to lead to a better correspondence and for a low order approximation the difference is negligible here.

Let us first calculate the lowest order: $c^y_2$ and $c^z_2$ for the $(3x^2-1)/2$ polynomial. As presented in Fig. \ref{spine}, we can rotate around the $x$ axis to make $c^z_2$ vanish. For this purpose, we can choose $(0,c^y_2,c^z_2)/c$ and orthogonal $(0,-c^z_2,c^y_2)/c$ as new $y$, $z$ directions, where $c=\sqrt{c^y_2+c^y_2}$. Thanks of it, in these coordinates the $c^y_2=c\geq0$, $c^z_2=0$. In other words, the molecule is '$\cup$'-like in $xy$ plane and nearly flat in $xz$ plane.

\begin{figure}[t!]
    \centering
        \includegraphics[width=8cm]{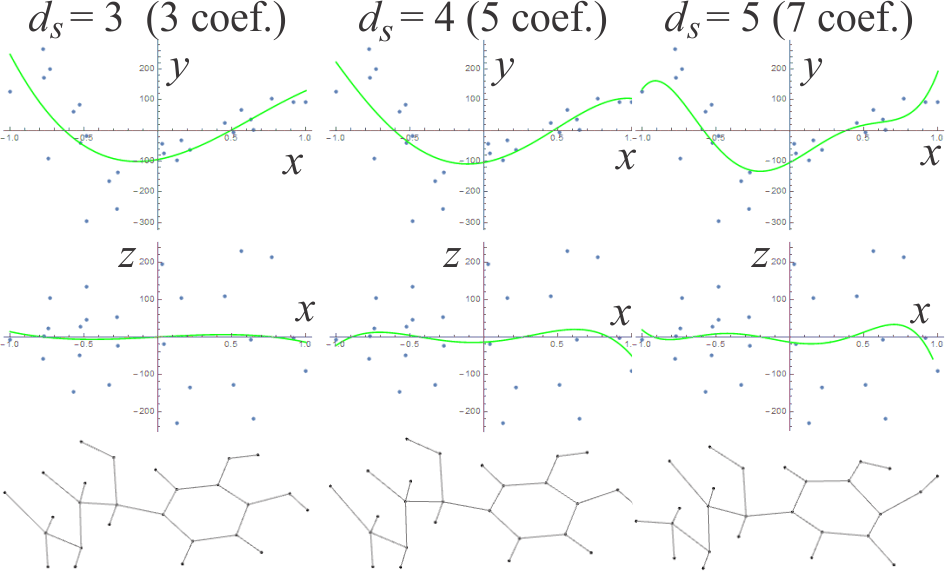}
        \caption{Figure \ref{spine} presented $d_s=2$ order spine, which required only 1 coefficient due to the choice of rotation. Here are presented spines of higher order, each of them require 2 additional coefficients: for $y$ and for $z$. The bottom row contains corresponding unbent molecules. As we can see, these additional coefficient does not seem important for epinephrine due to having '$\cup$' type of shape. However, they might be crucial for example for molecules of '$\sim$' type of shape.}
        \label{spine1}
\end{figure}

After this additional normalization step, we can eventually find some higher order coefficients $\{c_k^y, c_k^z\}$ for $k=2,\ldots,d_s$, where $d_s$ is the degree of approximation of spine and $c_2^z=0$. As presented in Fig. \ref{spine1}, $d_s=2$ should be sufficient for simple '$\cup$'-like molecules. However, higher order might be useful for example to handle '$\sim$'-like molecules.

For the final step we can unbend the molecule according to the approximation:
$$Y_i = y_i - \sum_{k=2}^{d_s} c_k^y P_k(x_i)\qquad Z_i = z_i - \sum_{k=2}^{d_s} c_k^z P_k(x_i)$$
These unbent molecules, chosen in ambiguous way for a given $d_s$, have still some complex shape around the $x$ axis - we will now discuss two different descriptors for them: BDC using ellipses as cross-sections, and BCH using cylindrical harmonics for this purpose. Having descriptor for unbent molecule, one should add the polynomials describing the bending to get representation of the final descriptor.

\begin{figure*}[t!]
    \centering
        \includegraphics[width=16cm]{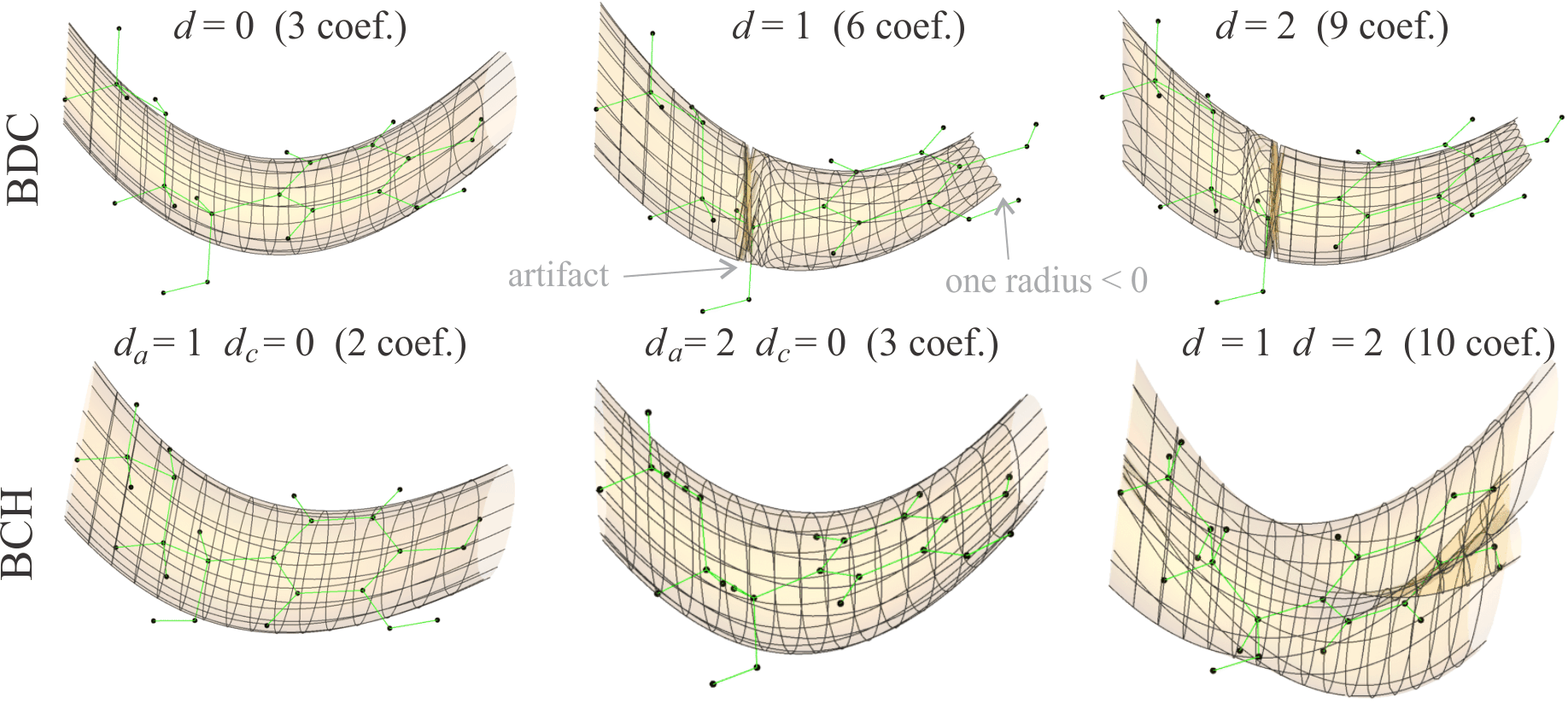}
        \caption{Some bent deformed cylinder (BDC) and bent cylindrical harmonics (BCH) representations of epinephrine for $d_s=2$ degree spine (1 additional coefficient describing bending, identical for all 6 cases). The BDC artifact is caused by changing order of eigenvalues (radii). While these representations are intended to close at $-1$ and $1$, the $d=1,2$ BDC closes earlier due to one radius going below 0. The $d_c=0$ BCH are circles in cross-sections.}
        \label{bent}
\end{figure*}
\subsection{Deformed cylinder descriptor (BDC)}
Let us start with describing the shape of the normalized and unbent molecule by ellipse shape evolving along $x$ axis, what can be imagined as a cylinder deformed in a complex way.

Fitting evolution of ellipse described by a few coefficients is a complex problem and various approaches can be chosen. Representations in Fig. \ref{bent} were obtained by fitting coefficients of $yz$ covariance matrix, what will be described now.

Covariance matrix consists in this case of average $Y_i^2$, $Z_i^2$ and $Y_i Z_i$ values. We can approximate evolution of these 3 averages as 3 polynomials of degree $d$ (generally various degrees can be chosen): fit degree $d$ polynomial to $(x_i,Y_i^2)$, $(x_i,Z_i^2)$ or $(x_i,Y_i Z_i)$ sets of points, getting $3(d+1)$ coefficients of our descriptor.

To decode shape described by them, like presented in Fig. \ref{bent}, we can start with analytical formula for eigenvectors and eigenvalues of $2\times 2$ matrix, getting a complex formula describing their evolution along $x$ axis by substitution of the fitted polynomials. The radii of ellipses in cross-section is given by square root of eigenvalue, its direction by corresponding eigenvector. It allows to define the presented parametric plots.

The analytical formula might switch the two eigenvalues, what leads to artifacts (additional 90 degree rotation) as emphasized in Fig. \ref{bent}. The polynomial approximation does not ensure positiveness of eigenvalues, what can lead to shortening of the cylinder from the expected $[-1,1]$ range, as presented in this figure.
\subsection{Cylindrical harmonics descriptor (BCH)}
\begin{figure}[t!]
    \centering
        \includegraphics[width=8cm]{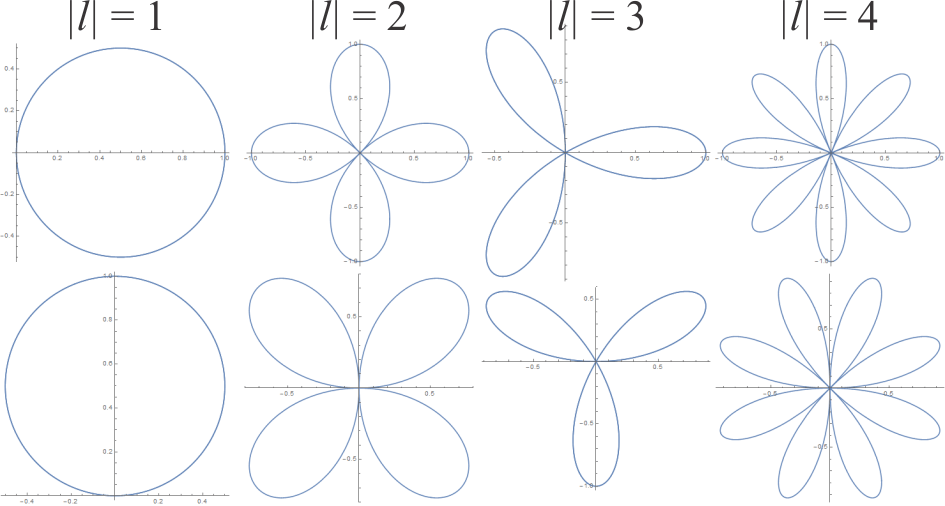}
        \caption{Polar plots of polar harmonics: $\cos(k\varphi)$ (upper) and $\sin(k\varphi)$ (lower).}
        \label{cyl}
\end{figure}

Alternative approach to describe a normalized unbent molecule are cylindrical harmonics: $r_l(\varphi)=\cos(l\varphi)$ for $l\geq 0$, $r_l(\varphi)=\sin(-l\varphi)$ for $l<0$. Some of them are presented in Fig. \ref{cyl}. We will restrict to $l=-d_c,\ldots, d_c$
for some degree of expansion $d_c$. The $d_c=0$ is means approximation with a circle - with radius evolving along $x$ axis in our case. The $|l|=1$ coefficients describe local direction of deformation. The $|l|=2$ can describe flattening. $|l|\geq 2$ have also some limited capability to represent forking of a molecule.

In our case we need to describe the evolution (along $x$ axis) of coefficients of these cylindrical harmonics - a simple way is to use separate degree $d_a$ polynomials for each of them. Finally, the number of coefficients is $(d_a+1)(2d_c+1)$.

To fit the coefficients of these polynomials, the points should be first transformed to $(x_i, r_i, \varphi_i)$ cylindrical coordinates. Then coefficients should be fitted for $P_k(x) \cos(l\varphi)$ base of functions (or $\sin$), where $P_k$ are Legendre polynomials. Analogously to the discussion in Section \ref{fitting}, it can be done by approximate projections, or by minimizing MSE.

\subsection{Describing distribution of mass and electron-negativity}
\begin{figure}[t!]
    \centering
        \includegraphics[width=8cm]{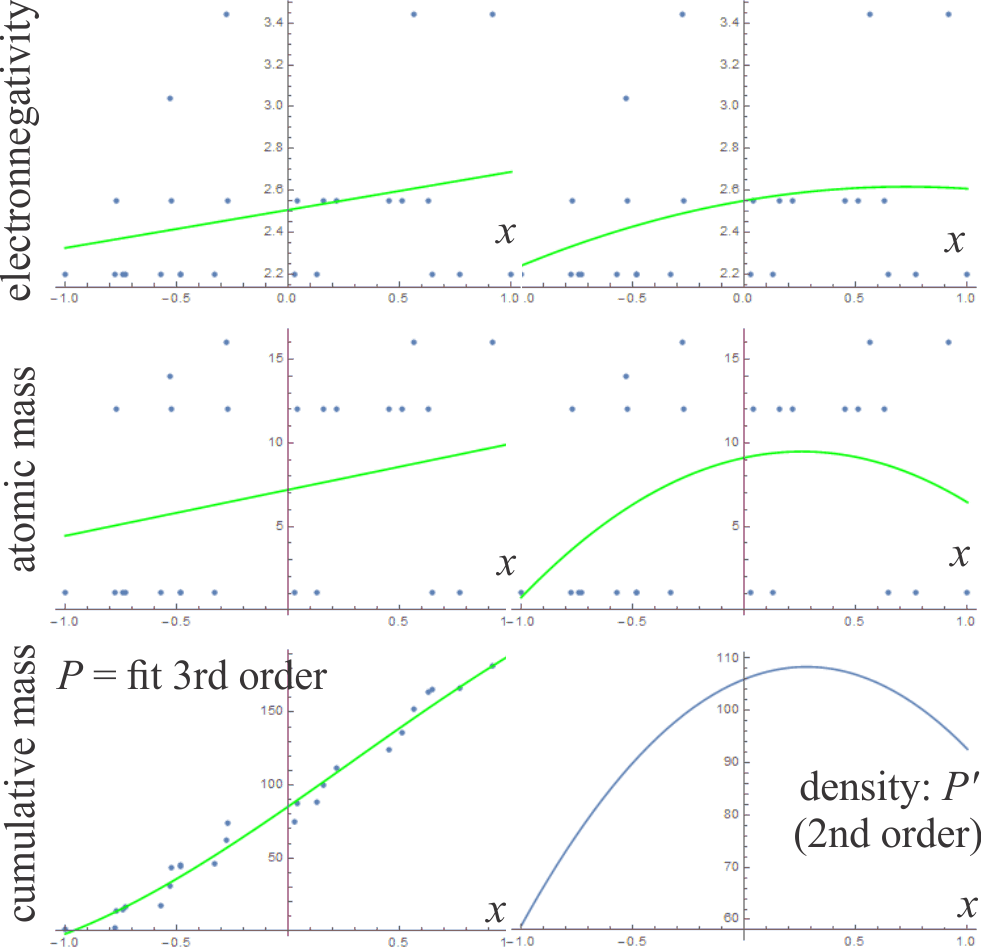}
        \caption{Approximation of electronnegativity and atomic mass distribution along the $x$ axis by using 1st and 2nd order polynomials - their coefficients can be used as additional coordinates of the descriptor. More appropriate for mass seems density descriptor: take cumulative mass up to given $x$, fit $d+1$ order polynomial, approximated density is its $d$-order derivative. The found maximum corresponds to the carbon ring. The integral of such fitted density should be close to the molecular mass.}
        \label{dist}
\end{figure}
The discussed descriptors produce some vector of real numbers describing the shape: for example 1 for width ($max-min$), at least one describing the bending (spine), and a few describing the evolution of cross-sections.

However, there are more properties missing in this description, which might provide some additional complementing information while evaluating resemblance. For example molecular mass, which should be correlated with the width of the molecule, or even better with volume of the bent cylinder. However, there can be fluctuations of densities of atoms inside such representation of the descriptor.

Another property, which is simple for quantitative evaluation, is electronegativity. It strongly influences activity around a given atom. Like presented in Fig. \ref{dist}, a simple way to describe its behavior for a molecule is to look at its evolution along the main axis $x$. It can be approximated by a polynomial, which coefficients can be used as additional coordinates of the descriptor vector. Its average value should be close to electronnegativeity of carbon (2.5) and so can be omitted. However, the linear coefficient carries important information about polarity of the molecule and so it might be beneficial to include it.

Similar fitting can be also applied for atomic weights along the main axis $x$ as presented in Fig. \ref{dist}. However, this simple approach does not necessarily correspond well to the density as it does not take atom frequencies into consideration. It can be repaired by also presented approach of cumulative mass, in analogy to the cumulative distribution function. Specifically, one should first replace atomic mass of a given atom with sum all masses on its left hand side (smaller or equal $x$). Then a polynomial should be fitted - its derivative approximates the density. We can use coefficients of this derivative as additional coordinates of our descriptor.

This approach of describing evolution along the main axis $x$ could be also used for different properties, for example for local ability to change conformation, what is essential while binding to a protein.

\section{Conclusions and further perspectives}
This article introduces framework of normalized rotation shape descriptors: based on emphasized both point and a specific rotation. PCA-normalized descriptor based on spherical harmonics is described first. Then there are proposed descriptors specialized for elongated shape of ligands, which are important purpose of virtual screening. The central bent spine of the molecule is approximated first, then evolution of cross-section along the main axis is described: by using ellipses (BDC) of cylindrical harmonics (BCH).

The general tools of this framework could be used to construct descriptors optimized for various applications, for example to get the best agreement with a reference metric on a set of molecules, which is representative for a given application. This reference metric should be appropriate for a given task and can be expensive, for example searching for the best alignment.
After calculating metric between all pairs in this set, the parameters of descriptor should be optimized:
\begin{itemize}
  \item which descriptor to choose - for example SH for globular molecules, BDC for elongated molecules, BCH if they can fork,
  \item choose interpolation orders, projection or minimization of MSE,
  \item choose normalization of coordinates (importance), maybe perform PCA for them to ensure independence, 
  \item design a metric between descriptors - taking into consideration their different natures, for example of width and coefficients of a polynomial,
  \item choose additional coordinates, for example describing distribution of electronnegativity or weight,
  \item decide if we are working close to symmetric situations, when additional mechanisms to ensure continuity should be applied.
\end{itemize}
As these descriptors operate on a single conformation, there should be generated separate descriptors for various conformation, for example differing by an angle step. As they usually should be similar, it might be beneficial to build descriptor from the average coefficients of such conformation ensemble, and add coordinates describing variance of these coefficients in the ensemble to describe the internal mobility of the molecule.

This article only outlines the basic framework - a lot of work needs to be done to optimize it for various applications, finding additional features which might be included in the descriptor, finding ways to construct descriptors: space of representations it operates on, ways of fitting them. For example more complex molecules can be forked, hence we should be able to describe disconnected cross-sections. Even more important is taking into consideration the possible changes of conformation, what is crucial while the binding process.

Another application is lossy compression, where additionally we should choose quantizers for all coefficients. The rate-distortion relation might be improved by including more coefficients, but using a more coarse quantization. For this kind of applications continuity might not be required.

\bibliographystyle{IEEEtran}
\bibliography{cites}

\begin{thebibliography}{1}
\providecommand{\url}[1]{#1}
\csname url@samestyle\endcsname
\providecommand{\newblock}{\relax}
\providecommand{\bibinfo}[2]{#2}
\providecommand{\BIBentrySTDinterwordspacing}{\spaceskip=0pt\relax}
\providecommand{\BIBentryALTinterwordstretchfactor}{4}
\providecommand{\BIBentryALTinterwordspacing}{\spaceskip=\fontdimen2\font plus
\BIBentryALTinterwordstretchfactor\fontdimen3\font minus
  \fontdimen4\font\relax}
\providecommand{\BIBforeignlanguage}[2]{{%
\expandafter\ifx\csname l@#1\endcsname\relax
\typeout{** WARNING: IEEEtran.bst: No hyphenation pattern has been}%
\typeout{** loaded for the language `#1'. Using the pattern for}%
\typeout{** the default language instead.}%
\else
\language=\csname l@#1\endcsname
\fi
#2}}
\providecommand{\BIBdecl}{\relax}
\BIBdecl

\bibitem{align}
C.~Lemmen and T.~Lengauer, ``Computational methods for the structural alignment
  of molecules,'' \emph{Journal of Computer-Aided Molecular Design}, vol.~14,
  no.~3, pp. 215--232, 2000.

\bibitem{kernel}
K.~M. Borgwardt, C.~S. Ong, S.~Sch{\"o}nauer, S.~Vishwanathan, A.~J. Smola, and
  H.-P. Kriegel, ``Protein function prediction via graph kernels,''
  \emph{Bioinformatics}, vol.~21, no. suppl 1, pp. i47--i56, 2005.

\bibitem{finger}
L.~Hodes, ``Selection of descriptors according to discrimination and
  redundancy. application to chemical structure searching,'' \emph{Journal of
  chemical information and computer sciences}, vol.~16, no.~2, pp. 88--93,
  1976.

\bibitem{USR}
P.~J. Ballester and W.~G. Richards, ``Ultrafast shape recognition for
  similarity search in molecular databases,'' in \emph{Proceedings of the Royal
  Society of London A: Mathematical, Physical and Engineering Sciences}, vol.
  463, no. 2081.\hskip 1em plus 0.5em minus 0.4em\relax The Royal Society,
  2007, pp. 1307--1321.

\bibitem{harm}
L.~Mavridis, B.~D. Hudson, and D.~W. Ritchie, ``Toward high throughput 3d
  virtual screening using spherical harmonic surface representations,''
  \emph{Journal of chemical information and modeling}, vol.~47, no.~5, pp.
  1787--1796, 2007.

\bibitem{PCAharm}
D.~Saupe and D.~V. Vrani{\'c}, \emph{3D model retrieval with spherical
  harmonics and moments}.\hskip 1em plus 0.5em minus 0.4em\relax Springer,
  2001.

\end{thebibliography}
\end{document}